\documentstyle[amssymb,preprint,aps]{revtex}

\renewcommand{\a}{a}

\begin{document}
\draft
\title{Using Markov chain Monte Carlo methods for estimating parameters with
gravitational radiation data.}
\author{Nelson Christensen$^{1}$ and Renate Meyer$^{2}$ \cite{email}}
\address{$^{1}$Physics and Astronomy, Carleton College, Northfield, Minnesota, 55057,%
\\
USA\\
$^{2}$ Department of Statistics, The University of Auckland, Auckland, New\\
Zealand}
\date{\today}
\maketitle

\begin{abstract}
We present a Bayesian approach to the problem of determining parameters for
coalescing binary systems observed with laser interferometric detectors. By
applying a Markov Chain Monte Carlo (MCMC)\ algorithm, specifically the
Gibbs sampler, we demonstrate the potential that MCMC techniques may hold
for the computation of posterior distributions of parameters of the binary
system that created the gravity radiation signal. We describe the use of \
the Gibbs sampler method, and present examples whereby signals are detected
and analyzed from within noisy data.
\end{abstract}



\narrowtext

\section{Introduction}

A number of collaborations around the world will be operating laser
interferometric gravitation radiation antennas within the next few years. In
the United States the Laser Interferometric Gravitational Wave Observatory
(LIGO) is soon to be operational, with 4 km arm length interferometers in
Hanford, Washington, and Livingston, Louisiana \cite{abra92}. A similar
French-Italian detector will be built in Europe (VIRGO) \cite{Brad90,Caron96}%
.

Coalescing binaries containing neutron stars (NS) or black holes (BH) are
likely to be the cleanest and most promising source of detectable radiation 
\cite{Thorne87}. Ultimately the LIGO-VIRGO network may observe binaries out
to a distance of 2 Gpc \cite{Finn93}. The detection of coalescing binary
events will provide physicists with extremely useful cosmological
information. Initially Schutz \cite{Schu86} noted that a detected signal
contains enough information to decipher the absolute distance to the system,
and hence the determination of the Hubble constant would be achieved through
the observed distribution of several binaries. Subsequent work \cite{Cutl94}
indicates that the uncertainty in the measured distance can be comparable to
the distance itself, but important cosmological tests will still be possible
through the observation of numerous mergers \cite{Finn96}.

In addition to the cosmological importance, accurate parameter estimation in
the observed coalescing binaries will provide a host of information of great
physical significance. Observation of the time of tidal disruption of an NS
- NS binary system may permit a determination of the NS radii and
information on the NS equation of state \cite{Cutl93}. The characteristics
of radiation in the post-Newtonian regime will provide insight into highly
non-linear general relativistic effects \cite{Cutl94,Flan98a,Flan98b}. The
formation of a BH at the end of a NS-NS coalescence, or the merger of two
BHs, will produce gravitational radiation as the system decays to a Kerr BH;
this is an extremely interesting radiation production regime \cite
{Flan98a,Flan98b}.

Application of Bayes' theorem is well suited to astrophysical observations 
\cite{Greg92}. The Bayesian versus frequentist approaches to gravitational
radiation data analysis are well presented by \cite{Finn97}. Parameter
estimation from the gravity wave signals of coalescing compact binaries
provides an important application of Bayesian methods \cite
{Finn93,Cutl94,Finn92,Nich98}. Difficulties with the calculation of Bayesian
posterior distributions have been overcome by the rapid development of
Markov Chain Monte Carlo (MCMC) methods in the last decade (see \cite{gilk96}
for an introduction). Although the initial MCMC algorithm dates back to \cite
{metr53}, the enormous potential that MCMC methods might hold for Bayesian
posterior computations remained largely unrecognized within the statistical
community until the seminal paper by Geman and Geman \cite{Gema84} in the
context of digital image analysis. Since then, MCMC methods have had a huge
impact on many areas of applied statistics. It has now become practical to
apply Bayesian methods to complex problems. Thus, we expect a similar effect
on gravitational wave data analysis. 

The initial goal of our research effort, presented in this paper, is to
demonstrate the usefulness of MCMC techniques for estimating parameters from
coalescing binary signals detected by laser interferometric antennas. The
Gibbs sampler \cite{gilk96} is one of the simpler MCMC\ techniques, and we
use it as a starting point for our investigation primarily because there is
readily available software \cite{BUGS96}. Our study of gravity wave signals
is conducted to 2.5 post-Newtonian (PN) order. The signals depend on five
independent parameters; the masses of the two compact objects, the amplitude
of the detected signal, the coalescence time and the phase of the signal at
coalescence. The Bayesian techniques we employ will not only give point
estimates of these parameters, but also produce their complete posterior
probability distribution that can be employed to summarize the uncertainty
of parameter estimates through posterior credibility intervals, for
instance. In contrast to frequentist confidence intervals, these do not rely
on large sample asymptotics and have a simple, natural interpretation.

The paper is organized as follows: In Section II we briefly review Bayesian
inference and describe the MCMC\ simulation technique we use, specifically
the Gibbs sampler and software for its implementation. In Section III we
present two examples where we use our MCMC\ approach to identify the
parameters which created the signal that is buried in synthesized LIGO\
noise. In Section IV we analyze a number of issues that will effect the
efficiency and calculational time of a MCMC approach to the coalescing
binary parameter estimation problem. We conclude with a discussion of our
results and the direction of future efforts in Section V.

\section{Bayesian Inference and Posterior Computation}

We briefly review the Bayesian approach to parameter estimation. Let us
assume the data consists of $n$ observations, ${\bf z}=(z_{1},\ldots ,z_{n})$%
, with joint PDF denoted by $p({\bf z}|\mbox{\boldmath $\theta$})$
conditional on unobserved parameters $\mbox{\boldmath $\theta$}=(\theta
_{1},\ldots ,\theta _{d})$. The PDF $p({\bf z}|\mbox{\boldmath $\theta$})$
is usually referred to as the {\em likelihood} and regarded as a function of 
$\mbox{\boldmath $\theta$}$. In contrast to the frequentist approach where $%
\mbox{\boldmath $\theta$}$ is regarded as fixed but unknown, the Bayesian
approach treats $\mbox{\boldmath $\theta$}$ as a random variable with a
probability distribution that reflects the researcher's uncertainty about
the parameters. Bayesian inference requires the specification of a prior PDF
for $\mbox{\boldmath $\theta$}$, $p(\mbox{\boldmath $\theta$})$, that should
take all information into account that is known about $%
\mbox{\boldmath
$\theta$}$ before observing the data. All information about $%
\mbox{\boldmath
$\theta$}$ that stems from the experiment should be contained in the
likelihood. Bayesian inference then answers the question: ``How should the
data ${\bf z}$ change the researcher's knowledge about $%
\mbox{\boldmath
$\theta$}$?'' Via an application of Bayes' theorem, by conditioning on the
known observations, this post-experimental knowledge about $%
\mbox{\boldmath
$\theta$}$ is expressed through the {\em posterior} PDF 
\begin{equation}
p(\mbox{\boldmath $\theta$}|{\bf z})=\frac{p(\mbox{\boldmath $\theta$})p(%
{\bf z}|\mbox{\boldmath $\theta$})}{m({\bf z})}\propto p(%
\mbox{\boldmath
$\theta$})p({\bf z}|\mbox{\boldmath 
$\theta$})
\end{equation}
where $m({\bf z})=\int p({\bf z}|\mbox{\boldmath $\theta$})p(%
\mbox{\boldmath 
$\theta$})d\mbox{\boldmath $\theta$}$ is the marginal PDF of ${\bf z}$ which
can be regarded as a normalizing constant as it is independent of $%
\mbox{\boldmath $\theta$}$. The posterior PDF is thus proportional to the
product of prior and likelihood.

The standard Bayesian point estimate of a single parameter, say $\theta_i$,
is the posterior mean 
\begin{equation}
\hat{\theta_i}= \int \theta_i p(\theta_i|{\bf z})d\theta_i
\end{equation}
where 
\begin{equation}
p(\theta _{i}|{\bf z})=\int \ldots \int p(\mbox{\boldmath $\theta$}|{\bf z}%
)d\theta _{1}\ldots d\theta _{i-1}d\theta _{i+1}\ldots d\theta _{d}.
\end{equation}
is the marginal posterior PDF obtained by integrating the joint posterior
PDF over all other components of $\mbox{\boldmath $\theta$}$ except $%
\theta_i $. A measure of the uncertainty of this estimate is the posterior
standard deviation or a 95\% credibility interval that contains the
parameter $\theta_i$ with 95\% probability, its lower and upper bound being
specified by the 2.5\% and 97.5\% percentile of $p(\theta_i|{\bf z})$,
respectively. Alternatives to the posterior mean are the posterior mode, aka
maximum a posteriori estimate (MAP), and the more robust posterior median.

As seen from equations (2) and (3), the calculation of posterior means
requires $d$-dimensional integration, one of the main issues that has made
the application of Bayesian inference so difficult in the past. This hurdle
has been overcome by the great advances in simulation-based integration
techniques, so called MCMC methods 
\cite{gilk96,chri98}. In MCMC, a Markov chain is constructed with the joint
posterior as its equilibrium distribution. Thus, after running the Markov
chain for a certain ''burn-in'' period, one obtains (correlated) samples
from the limiting distribution, provided that the Markov chain has reached
convergence. One popular construction principle is the Gibbs sampler, a
specific MCMC method that samples iteratively from each of the univariate
full {\em conditional} posterior distributions 
\begin{equation}
p(\theta _{i}|{\bf z},\theta _{1},\ldots ,\theta _{i-1},\theta _{i+1},\ldots
,\theta _{d}).
\end{equation}

Given an arbitrary set of starting values $\theta _{1}^{(0)},\ldots ,\theta
_{d}^{(0)}$ the algorithm proceeds as follows: 
\begin{eqnarray}
\mbox{simulate }\theta _{1}^{(1)} &\sim &p(\theta _{1}|{\bf z},\theta
_{2}^{(0)},\ldots ,\theta _{d}^{(0)})  \nonumber  \label{gibbs} \\
\mbox{simulate }\theta _{2}^{(1)} &\sim &p(\theta _{2}|{\bf z},\theta
_{1}^{(1)},\theta _{3}^{(0)},\ldots ,\theta _{d}^{(0)}) \\
&\vdots &  \nonumber \\
\mbox{simulate }\theta _{d}^{(1)} &\sim &p(\theta _{d}|{\bf z},\theta
_{1}^{(1)}\ldots ,\theta _{d-1}^{(1)})  \nonumber \\
&&  \nonumber
\end{eqnarray}
and yields $\mbox{\boldmath $\theta$}^{(m)}=(\theta _{1}^{(m)},\ldots
,\theta _{d}^{(m)})$ after $m$ such cycles. This defines a Markov chain 
that converges to the joint posterior as its equilibrium distribution \cite
{gilk96}. Consequently, if all the full conditional posterior distributions
are available, all that is required is sampling iteratively from these.
Thereby, the problem of sampling from an $d$-variate PDF is reduced to
sampling from $d$ univariate PDFs.

In many applications where the prior PDF is conjugate to the likelihood, the
full conditionals in fact reduce analytically to closed-form PDFs and we can
use highly efficient special purpose Monte Carlo methods for generating from
these (see e.g.\ \cite{devr86}). In general, however, we need a fast and
efficient black-box method to sample from an arbitrarily complex full
conditional posterior distribution in each cyclic step of the Gibbs sampler.
Such an all-purpose algorithm, so-called {\em adaptive rejection sampling}
(ARS) was developed by Gilks and Wild \cite{gilk92} for the rich class of
distributions with {\em log-concave} densities. We can use a recently
developed ``Metropolized'' version of adaptive rejection sampling (ARMS) for
non-logconcave distributions \cite{metr53,gilk95}. C-subroutines of ARS and
ARMS are available \cite{gilk95} and can thus be tailored to the full
conditional posteriors of the problem at hand.

Significant progress has been made in facilitating the routine
implementation of the Gibbs sampler with the help of BUGS (Bayesian
Inference Using Gibbs Sampling), a recently developed software package \cite
{BUGS96} by the Medical Research Council Biostatistics Unit, Institute of
Public Health, Cambridge, England. BUGS samples from the joint posterior
distribution by using the Gibbs sampler. For reviews on BUGS the reader is
referred to \cite{thom92,gilk94,gent97}. BUGS is available free of charge
from \smallskip
\begin{verbatim}
http://www.mrc-bsu.cam.ac.uk/bugs/Welcome.html
\end{verbatim}

BUGS can handle the two main tasks necessary for implementation of the Gibbs
sampler. These tasks are to i) construct and ii) to sample from the full
conditional posterior densities. 
Only the prior and sampling distributions for unobservables and observables,
respectively, have to be specified in a BUGS program. The tedious task of
constructing the full conditionals is automated by BUGS using directed
acyclic graphs \cite{meye00}. Sophisticated routines such as adaptive
rejection sampling to sample from log-concave full conditionals and MH
algorithms based on slice sampling to sample from non-logconcave full
conditional densities have been implemented in BUGS and are continuously
being refined. Furthermore, various methods to assess convergence, i.e.\
methods used for establishing whether an MCMC algorithm has converged and
whether its output can be regarded as samples from the target distribution
of the Markov chain, have been developed and implemented in CODA \cite
{coda95}. CODA is a menu-driven collection of SPLUS functions for analyzing
the output obtained by BUGS. Besides trace plots and the usual tests for
convergence, CODA calculates statistical summaries of the posterior
distributions and kernel density estimates. CODA is being maintained and
distributed by the same research group responsible for BUGS.

\section{Examples of Coalescing Binary Signals Detected by a Laser
Interferometer}

Our initial goal, presented in this paper, is to demonstrate the usefulness
of MCMC techniques for estimating parameters from coalescing binary signals
detected by laser interferometric antennas. We generated the coalescing
binary gravity wave signals to 2.5 PN order in both the time \cite
{Blanchet96,Blanchet96b} and frequency \cite{Tanaka00} domains. Noise that
simulates the LIGO II environment was synthesized in the time domain using
software from Finn and Daw \cite{Finn00}. The power spectral density of the
noise was calculated from long-time noise signals\ generated by \cite{Finn00}%
. The 2.5 PN frequency domain signals also served as the {\it templates} for
our extraction of the event parameters.

The total output registered by the detector, $z\left( t\right) $, is the sum
of the gravity wave signal, $s(t,\mbox{\boldmath $\theta$})$, that depends
on unknown parameters $\mbox{\boldmath $\theta$}$, and the noise $n(t)$,
namely $z\left( t\right) =s\left( t,\mbox{\boldmath $\theta$}\right)
+n\left( t\right) $ for $t\in \lbrack 0,t_{u}]$. We assume that the noise is
Gaussian with mean zero and known one-sided power spectral density $S_{n}(f)$%
. The signal-to-noise ratio ($SNR$) of the {\it detected} signal is $SNR=%
\sqrt{2\int_{-\infty }^{\infty }\frac{\tilde{s}\left( f\right) \tilde{s}%
\left( f\right) }{S_{n}\left( f\right) }df}$ with $\tilde{s}\left( f\right)
=\int_{-\infty }^{\infty }s\left( t\right) e^{2\pi ift}dt$ the Fourier
transform of the function $s(t)$ \cite{Finn96}. The {\it likelihood} is
given by 
\begin{equation}
p({\bf z}|\mbox{\boldmath $\theta$})=K\exp \left[ 2\left\langle z,s(%
\mbox{\boldmath $\theta$})\right\rangle -\left\langle s(%
\mbox{\boldmath
$\theta$}),s(\mbox{\boldmath $\theta$})\right\rangle \right]
\end{equation}
where $K$\ is a constant and $\left\langle a,b\right\rangle =\int_{-\infty
}^{\infty }df\frac{\tilde{a}(f)\tilde{b}^{\ast }(f)}{S_{n}(f)}$ the inner
product of two functions $a,b$ \cite{Finn93}. 

Since $\tilde{z}(f)=\tilde{z}^{\ast }(-f)$ for real signals, we can express
the likelihood as 
\begin{equation}
p({\bf z}|\mbox{\boldmath $\theta$})=K\exp \left[ -2\int_{0}^{\infty }df%
\frac{Re\left\{ (\tilde{z}(f)-\tilde{s}(f,\mbox{\boldmath $\theta$}))(\tilde{%
z}(f)-\tilde{s}(f,\mbox{\boldmath $\theta$}))^{\ast }\right\} }{S_{n}(f)}%
\right] .
\end{equation}
For discretized data the likelihood takes the form 
\begin{equation}
p({\bf z}|\mbox{\boldmath $\theta$})=K\ast \exp \left[ -2%
\sum_{i=i_{l}}^{i_{u}}\frac{Re\left( \tilde{z}(i\ast \Delta f)-\tilde{s}%
(i\ast \Delta f,\mbox{\boldmath $\theta$})\right) \left( \tilde{z}(i\ast
\Delta f)-\tilde{s}(i\ast \Delta f,\mbox{\boldmath $\theta$})\right) ^{\ast }%
}{S_{n}\left( i\ast \Delta f\right) }\right] 
\end{equation}
where $i_{l}\ast \Delta f$ and $i_{u}\ast \Delta f$ correspond to the lower
and upper limits of the frequency range examined and $\Delta f$ is the
resolution of the discretized domain data.

As detailed in \cite{Tanaka00}, the signal depends on five parameters,
namely the masses $m_{1},m_{2}$ of the two compact objects, the coalescence
time $t_{c}$, the phase of the wave $\varphi _{0}$, and the amplitude $N$
through 
\begin{equation}
\tilde{s}(f,\mbox{\boldmath $\theta$})=Ne^{i\varphi _{0}}f^{-7/6}e^{i\left(
\psi \left( f\right) +2\pi ft_{c}\right) }
\end{equation}
with $\mbox{\boldmath $\theta$}=(m_{1},m_{2},t_{c},\varphi _{0},N)$ and 
\begin{equation}
\psi (f)=\sum_{i=1}^{5}\a_{i}\varsigma _{i}(f)
\end{equation}
where 
\begin{eqnarray}
a_{1} &=&\frac{3}{128\eta }q^{-5/3},  \label{template} \\
a_{2} &=&\frac{1}{384\eta }\left( \frac{3715}{84}+55\eta \right) q^{-1}, 
\nonumber \\
a_{3} &=&\frac{-1}{128\eta }48\pi q^{-2/3},  \nonumber \\
a_{4} &=&\frac{3}{128\eta }\left( \frac{15293365}{508032}+\frac{27145}{504}%
\eta +\frac{3085}{72}\eta ^{2}\right) q^{-1/3},  \nonumber \\
a_{5} &=&\frac{\pi }{128\eta }\left( \frac{38645}{252}+5\eta \right) , 
\nonumber \\
&&  \nonumber
\end{eqnarray}
and $\varsigma _{1}=f^{-5/3}$, $\varsigma _{2}(f)=f^{-1}$, $\varsigma
_{3}(f)=f^{-2/3}$, $\varsigma _{4}(f)=f^{-1/3}$, $\varsigma _{5}(f)=\ln (f)$%
, the total mass $m_{t}=m_{1}+m_{2}$, $q=\pi Gm_{t}/c^{3}$, and the mass
ratio $\eta =m_{1}m_{2}/m_{t}^{2}$.

The {\it templates} for the gravity wave signal, $\tilde{s}(f,%
\mbox{\boldmath $\theta$})$, were always generated in the frequency domain
to 2.5PN\ order according to the formulas given in \cite{Tanaka00} because
the entire likelihood computation is also done entirely in the frequency
domain. Examples of our Gibbs sampler code for use with BUGS can be obtained
electronically \cite{NLC00}.

\subsection{Results for Signals F in the Frequency Domain}

The published 2.5 PN\ order time domain templates are not quite the exact
Fourier transform of the 2.5 PN frequency domain templates, hence we felt
the necessity to test our method with signals we generated in both time or
frequency space. For signals generated in the frequency domain we used
published 2.5 PN order results%
\index{template} \cite{Tanaka00}. For this example we chose the masses of
the compact objects to be $m_{1}=1.4M_{\odot }$ and $m_{2}=3.5M_{\odot }$,
the coalescence time was $t_{c}=1ms$, and the phase of the wave of $\varphi
_{0}=0.123$. The amplitude of the wave was adjusted to create appropriate $%
SNR$ values. The noise was generated in the time domain \cite{Finn00}, and
subsequently transformed to the frequency domain.

We assumed noninformative {\it a priori} distributions for all of our
parameters; namely a uniform distribution on 0.3$M_{\odot }$ to 12$M_{\odot
} $ for each of the two compact objects ($m_{1}$ and $m_{2}$) and a uniform
distribution on $0$ to $50ms$ for $t_{c}$. The gravity wave phase will lie
between and $-\pi $ to $\pi $\ for $\varphi _{0}$, but we assume a uniform a
priori distribution between $-2\pi $ to $2\pi $ so that the converged chain
can more easily sample its region of interest. The amplitude of the incoming
gravity wave has a dependence of $%
\tilde{h}\left( f\right) =N\eta ^{1/2}m_{t}^{5/6}f^{-7/6}$. Our simulated
signal thereby has $m_{t}=4.9M_{\odot }$ and $\eta =0.2041$. We used a
uniform {\it a priori} for the amplitude term $N$ on the interval $-10^{-19}$
to $10^{-19}$. This was subsequently renormalized with a factor of $10^{25}$
for computational reasons yielding a uniform {\it a priori} distribution for 
$N$ on the interval $-10^{6}$ to $10^{6}$. We assumed that the compact
objects had no spin, and hence these parameters were not included in this
study.

Unique gravity wave signals do not depend on $m_{1}$ and $m_{2}$, but
instead on the total mass $m_{t}$ and the mass ratio $\eta $. Hence it is
the posterior probability distribution functions of $m_{t}$ and $\eta $ that
provide the best information about the system. The posterior distributions
of $m_{1}$ and $m_{2}$ from the MCMC are unnecessarily wide due to
oscillations of each chain between the two possible values. Estimates and
statistical properties of $m_{1}$ and $m_{2}$ must be inferred from the
distributions for $m_{t}$ and $\eta $.

In our initial implementation of the Gibbs sampler it was found that it
takes a prohibitively long time for the chain to {\em burn in} and sample
from the correct posterior distribution. Instead, an efficient procedure
that allowed the chain to more efficiently explore the phase space was one
that is analogous to {\it simulated annealing }\cite{Gilks96}. In this
procedure we use a likelihood of the form $L=K\exp \left[ 2\left\langle
z,s\right\rangle -\left\langle s,s\right\rangle \right] $ with $\left\langle
z,s\right\rangle =\int_{-\infty }^{\infty }df\frac{\tilde{z}(f)\tilde{s}%
^{\ast }(f)}{T\ast S_{n}(f)}$, $\left\langle s,s\right\rangle =\int_{-\infty
}^{\infty }df\frac{\tilde{s}(f)\tilde{s}^{\ast }(f)}{T\ast S_{n}(f)}$. The
auxiliary variable $T$ is a {\it pseudo-temperature}. If $T$ is chosen
large, $T>>1$, we have {\it heating} and the MCMC does not get trapped in
particular regions of phase space for too long. It essentially corresponds
to increasing the variance of the posterior distribution to allow for wider
jumps. Thus, the chain can reach all regions of the state space. In our
study we typically start with $T\symbol{126}500$ and allow the chain to
''burn-in'' and find equilibrium. For each value of $T $ the mean values for
the parameters are computed and used as the starting values for the next
chain with its reduced value of $T$. The $T$ term can be quickly brought to $%
T=1$ and the final kernel densities generated. Combining simulated annealing
with MCMC samplers has been demonstrated to improve the efficiency of chains
that mix poorly in their phase space \cite{Jennison93}.

When we analyzed the data in our MCMC program we had a frequency resolution
of $\Delta f=1Hz$ and only utilized the frequency range $30Hz-730Hz$; this
range was based on reasonable assumptions of LIGO\ performance. Reducing the
upper frequency limit does not greatly effect the performance of the MCMC\
results; this makes sense because for a coalescing binary signal most of the
power of the signal is at the lower frequencies. The largest amplitude
signal may happen at a high frequency, but the binary emits more cycles at
lower frequencies. Although we have not studied this issue in depth, it
appears that good MCMC\ performance will result even if one only includes
frequencies up to $\symbol{126}200Hz.$

In Figs. 1-3 we present the results of this part of our investigation. We
see that for large $SNRs$ we can accurately predict the parameters producing
the signal. At the lower $SNRs$ it takes the chain much longer to burn-in
and find the correct parameters, and when the $SNR$ is too low the chain
will not converge and no useful information will be discerned. The results
for $SNR=4.4$ are further illustrated in Fig. 4, where we can observe the
kernel densities for the parameters\ generated from the Gibbs sampler.

Fig. 5 displays the operation of the Gibbs sampler, whereby the trace plots
for the parameters are displayed. This is the result for $SNR=4.4$ when we
have set the annealing temperature to $T=100$. One can see how the chain 
{\it burns-in} and achieves convergence. We would let the chain run for 
\symbol{126}10,000 iterations, compute the mean values for the parameters
(excluding points prior to {\it burn-in}), and use the calculated mean
parameter values as initial conditions for a new Markov chain with a
diminished annealing temperature. Once we have a $T=1$ then 10,000
iterations typically produce an adequate and informative kernel density.

Extensive convergence diagnostics were calculated for all of the parameters
using the CODA software \cite{coda95}. All chains passed the Heideberger -
Welch stationary test. The Raferty - Lewis convergence diagnostics confirmed
the thinning and burn-in were sufficient. Lags and autocorrelations within
each chain were reasonably low. Geweke Z-scores were low for all parameters.
These convergence diagnostics are described in \cite{coda95} (and references
therein).

Continuing the use of the $SNR=4.4$ example, we can decipher the masses of
the individual compact objects. The result of the Gibbs sampler can give us
estimates of these parameters by using simple summary statistics like the
sample average and empirical percentiles. These yield posterior means of $%
m_{t}=4.891M_{\odot }\ $and $\eta =0.2048$, plus 2.5 \ to 97.5 percentile
ranges of $m_{t}=\left( 4.882\text{ to }4.898\right) M_{\odot }$ and $\eta
=0.2043$ to $0.2055$. This then implies, compact object masses of $%
m_{1}=\left( 3.485\pm 0.01\right) M_{\odot }$ and $m_{2}=\left( 1.406\pm
0.008\right) M_{\odot }$.

\subsection{Results for Signals Generated in the Time Domain}

When we generated signals in the time domain the parameters were also chosen
arbitrarily. The generated signals were made to 2.5 PN \cite
{Blanchet96,Blanchet96b}. The masses of the two compact object were $%
8M_{\odot }$ and $9M_{\odot }$, the angle of inclination of the orbit was $%
\iota =\pi /4$, and the orientation of the gravity wave source with respect
to the laser interferometer was $\varphi =2.2$, $\theta =1.1$, and $\psi
=3.3 $ (all in radians). In order to adjust the signal to noise ratio we
effectively changed the source to detector distance. The interferometer
noise was also computer generated \cite{Finn00}, and the signal and noise
were summed together. We created signals of $32s$ length, with $16384$ data
points per second. The temporal signal was Fourier transformed to the
frequency domain.

Templates for the likelihood were again 2.5 PN \cite{Tanaka00}. The results
below were based on a MCMC\ investigation that ranged from $30Hz-130Hz$,
with frequency resolution of $\Delta f=1Hz$. Increasing the upper frequency
did not affect the results, but only slowed the calculation. This is again
consistent with the fact that most signal power is at the lower frequency.
The choice of an upper frequency for the MCMC will effect the speed of the
calculation and ultimately its ability to accurately estimate parameters;
this is a topic we are currently investigating and will be the subject of a
future publication.

In Figs.\ 6-8 we present the results of this part of our investigation. We
see that for large $SNRs$ we can again accurately predict the parameters
producing the signal. The results for $SNR=4.4$ are presented in Fig.\ 9,
where we can observe the kernel densities for the parameters that we
generated from the Gibbs sampler. Fig.\ 10 displays the operation of the
Gibbs sampler, whereby the trace plots for the parameters are displayed.
This is the result for $SNR=4.4$ when we have set the annealing temperature
to $T=100$. One can see how the chain {\it burns-in} and converges.

\section{Analysis Issues}

There are a host of topics that need to be addressed before one could say
that MCMC\ techniques will be truly applicable and useful with LIGO data.
However, we have demonstrated that the Gibbs sampler does have potential
usefulness, and can successfully find the signal and make statistical
statements about the parameters. Numerous issues pertaining to MCMC use with
LIGO\ data are discussed below.

The speed at which the MCMC calculation runs on computers is a concern of
paramount importance. Numerous issues influence the speed of the
calculation. For the study presented in this paper, we would generate $32s$
of data in the time domain. The data corresponded to a sampling rate of $%
16384Hz$. Due to the character of the signal source and LIGO noise we only
considered signal frequencies above $30Hz$.

The choice of an upper frequency limit can significantly influence the speed
of the MCMC program. However, one can not arbitrarily reduce the upper
frequency too much or the ability to estimate parameters will degrade. If
our templates only consider the orbital parameters of the binary system, and
not the ringdown of the newly formed black hole then the largest useful
frequency will correspond to twice the instantaneous orbital frequency of
the last stable orbit before free-fall, $f=c^{3}/\left( 6^{3/2}\pi
Gm_{t}\right) $. In this study we only considered compact objects with
masses between $0.3M_{\odot }$ and $10M_{\odot }$. The frequency could
therefore range from as large $7300Hz$ for two $0.3M_{\odot }$ objects, to $%
1570Hz$ for two $1.4M_{\odot }$ neutron stars, to $220Hz$ for two $%
10M_{\odot }$ black holes. One should also remember that the Fourier
transform of the signal falls off like $\left| h\left( f\right) \right|
\varpropto f^{-7/6}$. Consequently the power of the signal is dominated by
low frequencies; the binary spends more time emitting relatively low
amplitude gravity waves at lower frequency as opposed to a shorter time
producing larger amplitude signals at higher frequencies. Establishing an
effective upper frequency choice will depend on the decision for the {\it a
priori} distribution of the masses. It will also depend on analyses that
investigate the behavior of the MCMC as the upper frequency varies. We
anticipate that effective parameter estimation studies of events will vary
the upper frequency and search for a convergence in behavior. As the upper
frequency of the data is diminished the speed of the calculation increases.

The frequency resolution of the data is another aspect that will affect
program speed and parameter estimation ability. For example, with $32s$ of
data the Fourier transform points are separated by $1/32$ $Hz$. We found
this precise frequency resolution to be unnecessary. We increased the
program speed without diminishing parameter estimation ability by decreasing
the resolution to $1Hz.$ All the results presented in this paper utilized a
frequency resolution of $1Hz$. However, we feel that a detailed study of the
coupling of frequency resolution with computational speed and effectiveness
will be necessary.

The {\it simulated annealing} procedure we used to adequately sample the
parameter space is another topic where studies will be needed\ in order to
optimize the speed of the routine. For the variables we used in our program,
either in terms of the compact masses $m_{1}$ and $m_{2}$ or the total mass $%
m_{t}$ and the mass ratio $\eta $, it was necessary to introduce an {\it %
pseudo-temperature} $T$ into the likelihood to initially {\it burn-in} the
chain and converge to the correct parameters. Without $T$ it would take a
prohibitively long time for the chain to converge; the incorporation of
simulated annealing into MCMC techniques has been demonstrated to decrease
the burn-in time for the chain \cite{Jennison93}. In the studies presented
here we would typically start with $T\symbol{126}500$, and it could take
anywhere from $10^{3}$ to $2\times 10^{4}$ cycles for the chain to reach
convergence. Smaller $SNR$ events took longer to {\it burn-in}.\ The chain
would run for about $10^{4}$ after burn-in, and the mean values for the
parameters were computed and used as the starting values for the next chain
with its reduced value of $T$. The value of $T$ would be decreased by an
order of magnitude, so that a typical procedure would involve runs with $%
=500,$ $50,$ $5$ and then $1$. This cooling schedule is definitely not
optimized for speed and efficiency. The use of better coordinates \cite
{Sathya00} may provide a more well behaved parameter space, which in turn
could help the chain mix more efficiently and reduce the time needed for
simulated annealing. This will be investigated.

All of the calculations presented in this paper were conducted on a $%
500\;MHz $ pc. When the frequency span of the study extended from $30Hz$ to $%
730Hz$, with $1Hz$ resolution, $10^{4}$ cycles of the MCMC took 1.5 hours.
We often needed about $2.5\times 10^{4}$ cycles to generate good kernel
densities. When the frequency span extended from $30Hz$ to $130Hz$, with $%
1Hz $ resolution, $10^{4}$ cycles of the MCMC typically took approximately
10 minutes.

\section{Discussion}

In this paper we have demonstrated that the Gibbs sampler can be used to
estimate the parameters for a coalescing binary system for signals detected
by LIGO antennas. While we have not yet optimized this procedure for speed,
we have shown that within just a few hours of running time on a $500MHz$ pc
we can generate kernel densities for the parameters. The MCMC can replace a
systematic march through a grid of templates in parameter space, and instead
make a probabilistic random walk that is helped by the weighting of the
product of the likelihood and the $a$ $priori$ distributions of the
parameters.

We have only concentrated on using our MCMC procedure for estimating signals
in data sets where a signal is assumed to exist. Using MCMC techniques as a
method to find signals from the continuous output of the LIGO detectors is
another research topic that is not covered here. A question that must be
answered is how efficiently can the MCMC\ method identify signals, and how
often does it miss them.

A great advantage of MCMC\ methods is that the calculational time does not
scale exponentially with parameter number, but in fact scales almost
linearly. Applications of state-space modelling in finance, such as
stochastic volatility models applied to time series of daily exchange rates
or returns of stock exchange indices, easily have 1000 - 5000 parameters;
specially tailored MCMC\ algorithms can effectively and efficiently sample
the phase space \cite{meye00,jacq94,meyu00,kim98}. The number of parameters
for coalescing binary signals will grow when, for example, the spin of the
compact objects is included. We will also eventually expand our MCMC study
to the problem of examining the signals detected by two or more
interferometers simultaneously. In addition to the parameters for pertaining
to the coalescing binary, there will be the delay in arrival times between
the detectors and the polarization sensitivities that can then infer the
location in the sky of the source \cite{Ande00}. MCMC\ methods offer great
promise for parameter estimation with coalescing binary signals, especially
as parameter numbers increase.

\acknowledgments 
This work was supported by the National Science Foundation Grant
PHY-0071327, Carleton College, The Royal Society of New Zealand Marsden
Fund, and the University of Auckland Research Committee.

\appendix

\appendix 

\newpage

\begin{description}
\item  {\Huge Figure Captions}
\end{description}

Fig. 1 Estimate of the total mass, $m_{t}$, versus $SNR$. The actual total
mass for the signal is $m_{t}=4.9M_{\odot }$. Error bars correspond to the $%
2.5$ and $97.5$ percentile of the posterior distribution of $m_{t}$ (which
gives a 95\% posterior credibility interval for $m_{t}$).

\bigskip

Fig. 2 Estimate of the mass ratio, $\eta $, versus $SNR$. The actual mass
ratio for the signal is $\eta =0.2041$. Error bars correspond to the $2.5$
and $97.5$ percentile of the posterior distribution of $\eta $ (which gives
a 95\% posterior credibility interval for $\eta $).

\bigskip

Fig. 3 Estimate of the (rescaled) amplitude of the gravity wave, $N$, versus 
$SNR$. The amplitude of the inferred signal varies linearly with the $SNR$,
as it should. Error bars correspond to the $2.5$ and $97.5$ percentile of
the posterior distribution of $N$ (which gives a 95\% posterior credibility
interval for $N$).

\bigskip

Fig. 4 The kernel densities for the parameters $\left( m_{t},\text{ }\eta ,%
\text{ }N,\text{ }t_{c},\text{ and }\varphi _{0}\right) $,\ generated from
the Gibbs sampler with $SNR=4.4.$

\bigskip

Fig. 5 Example of the operation of the Gibbs sampler, with trace plots for
the parameters displayed, with $SNR=4.4$ and the simulated annealing
pseudo-temperature of $T=100$. One can see how the chain {\it burns-in} and
achieves convergence.

\bigskip

Fig. 6 Estimate of the total mass, $m_{t}$, versus $SNR$. The actual total
mass for the signal is $m_{t}=17M_{\odot }$. Error bars correspond to the $%
2.5$ and $97.5$ percentile of the posterior distribution of $m_{t}$ (which
gives a 95\% posterior credibility interval for $m_{t}$).

\bigskip

Fig. 7 Estimate of the mass ratio, $\eta $, versus $SNR$. The actual mass
ratio for the signal is $\eta =0.2491$. Error bars correspond to the $2.5$
and $97.5$ percentile of the posterior distribution of $\eta $ (which gives
a 95\% posterior credibility interval for $\eta $).

\bigskip

Fig. 8 Estimate of the (rescaled) amplitude of the gravity wave, $N$, versus 
$SNR$. The amplitude of the inferred signal varies linearly with the $SNR$,
as it should. Error bars correspond to the $2.5$ and $97.5$ percentile of
the posterior distribution of $N$ (which gives a 95\% posterior credibility
interval for $N$).

\bigskip

Fig. 9 The kernel densities for the parameters $\left( m_{t},\text{ }\eta ,%
\text{ }N,\text{ }t_{c},\text{ and }\varphi _{0}\right) $,\ generated from
the Gibbs sampler with $SNR=4.4.$

\bigskip

Fig. 10 Example of the operation of the Gibbs sampler, with trace plots for
the parameters displayed, with $SNR=4.4$ and the simulated annealing
pseudo-temperature of $T=100$. One can see how the chain {\it burns-in} and
achieves convergence.

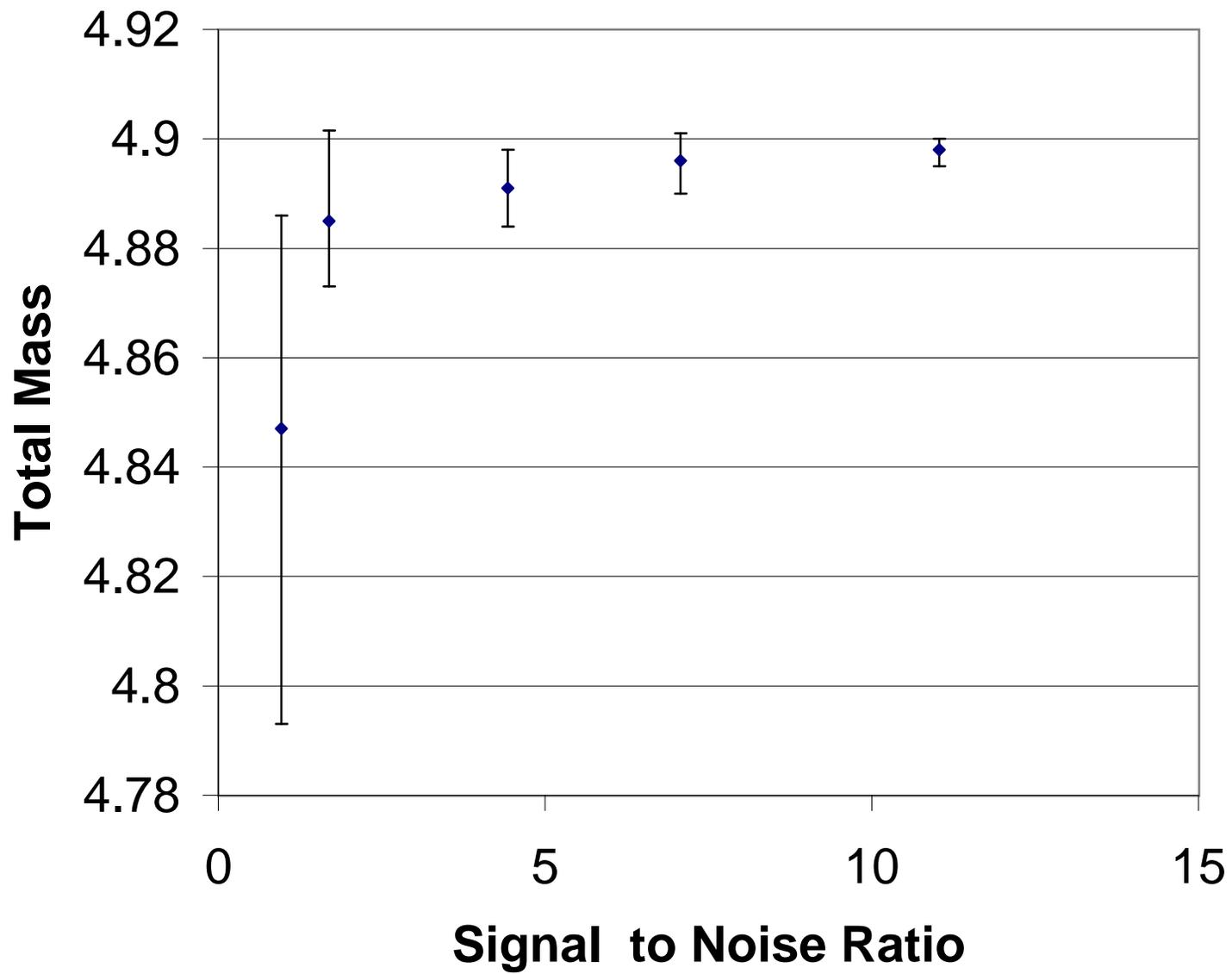
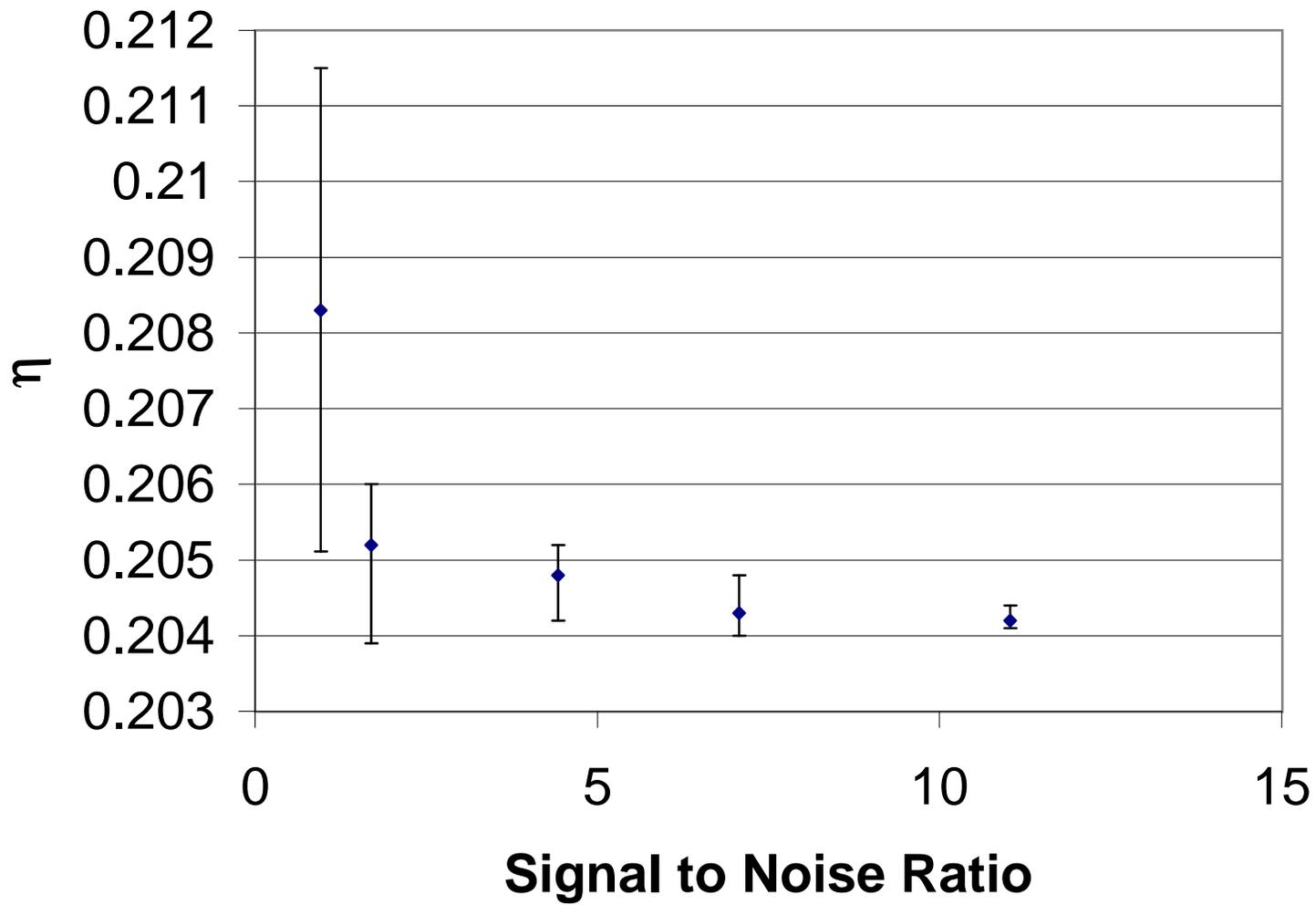
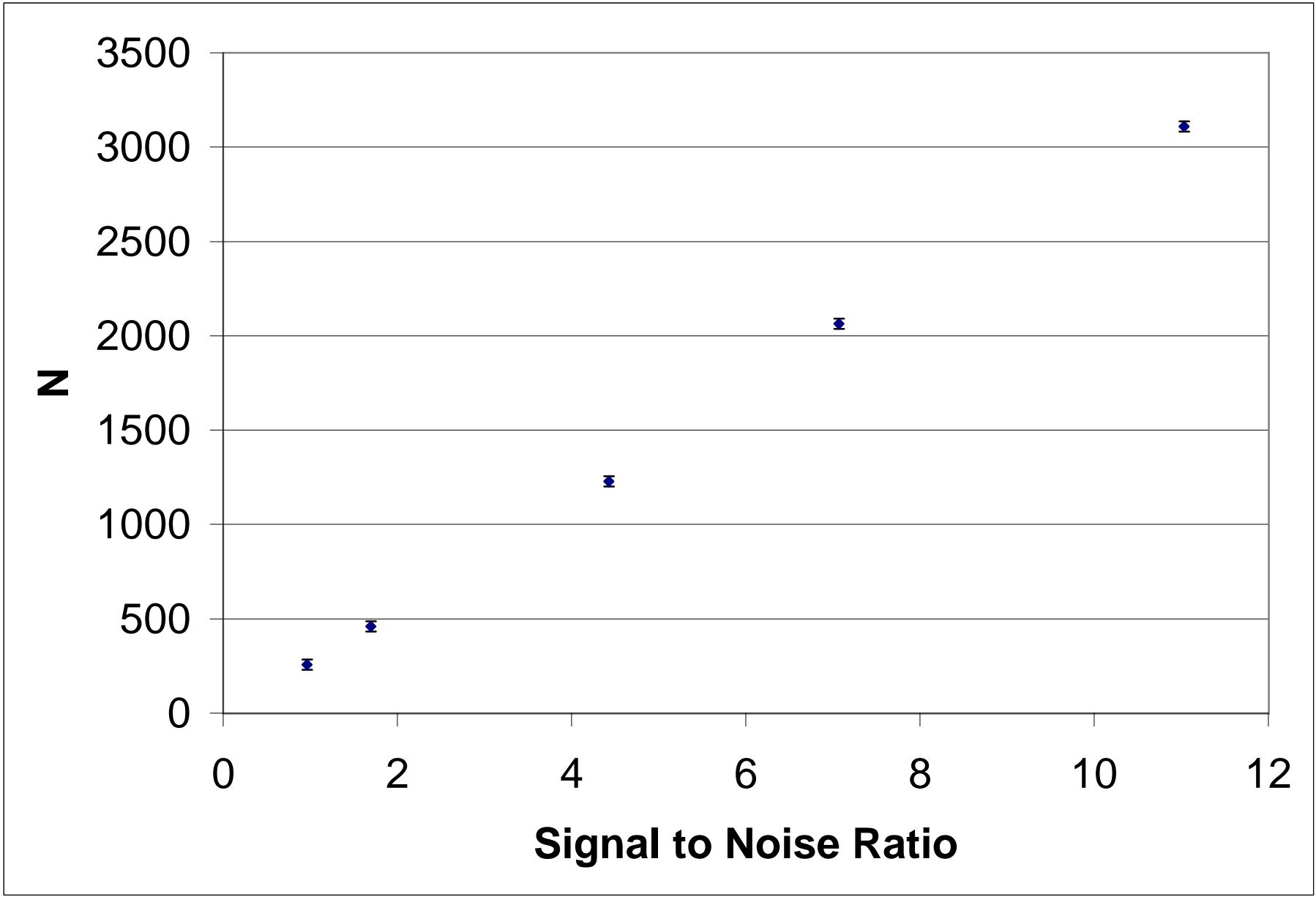
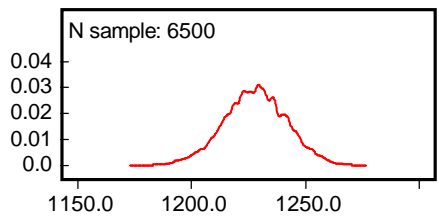
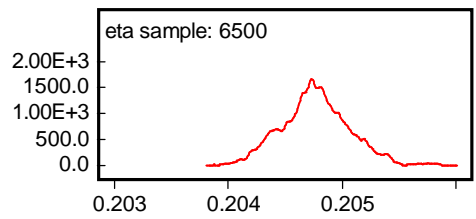
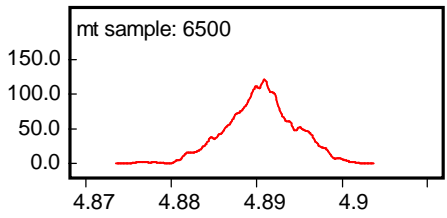
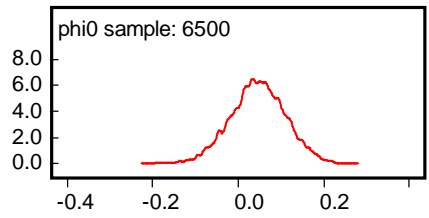
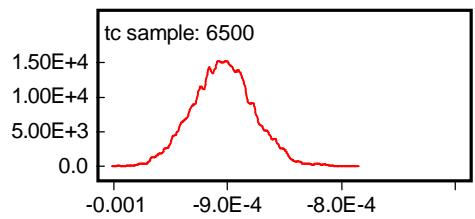
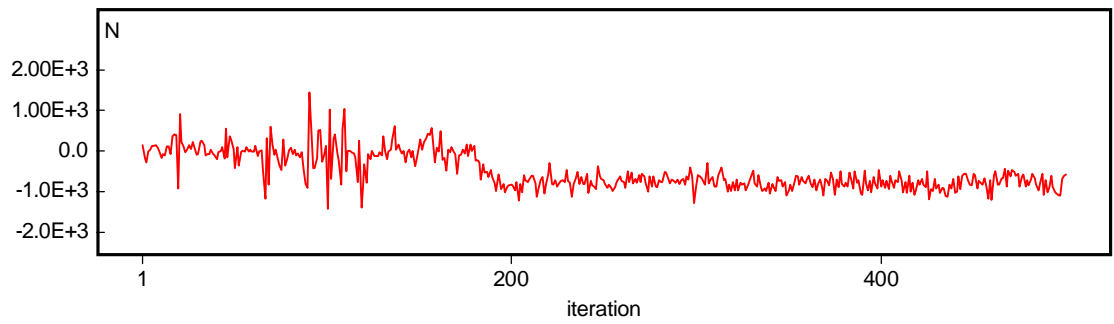
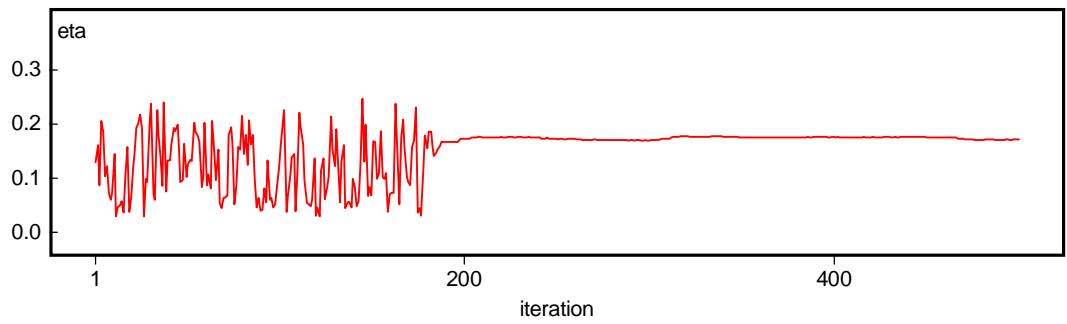
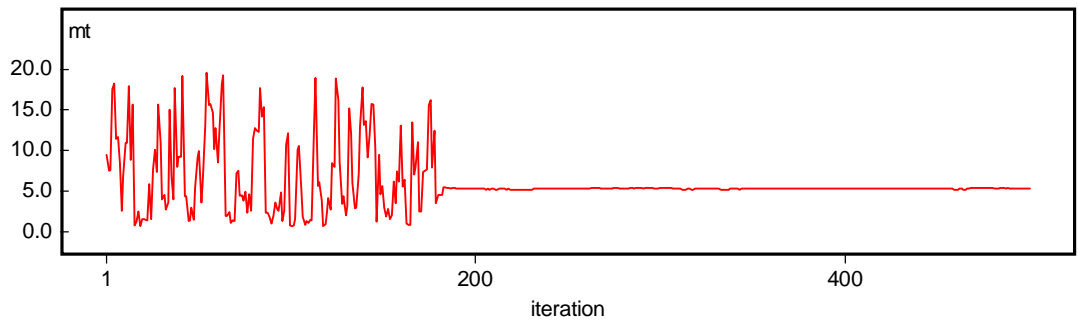
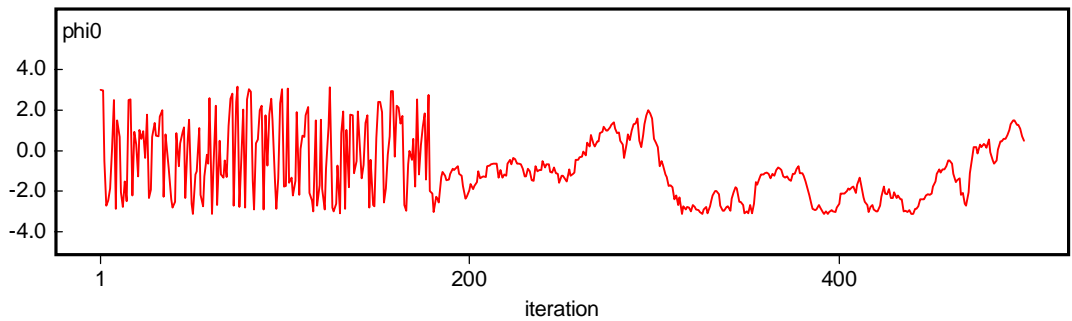
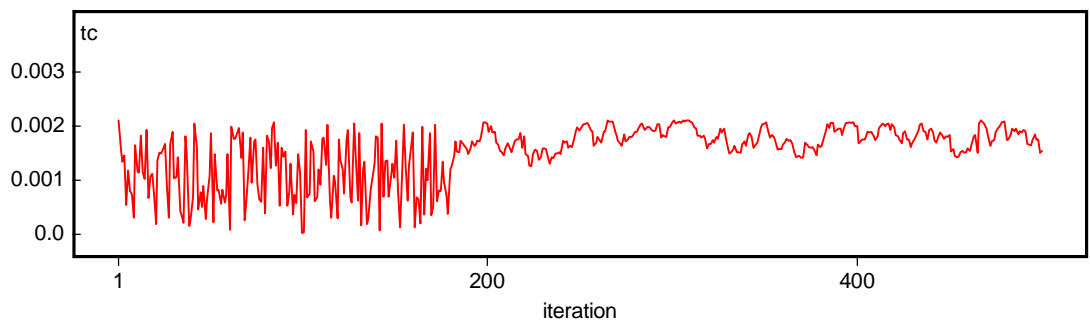
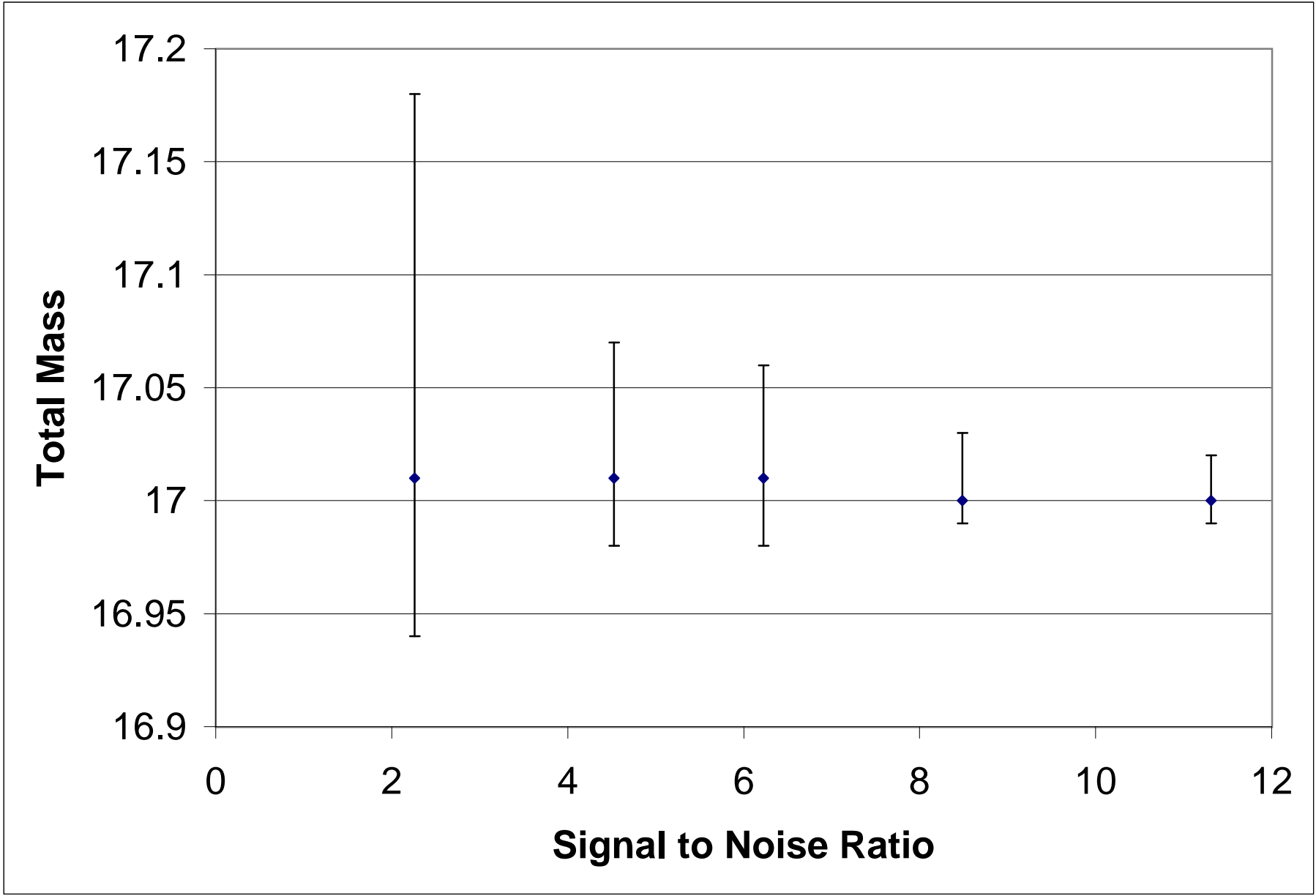
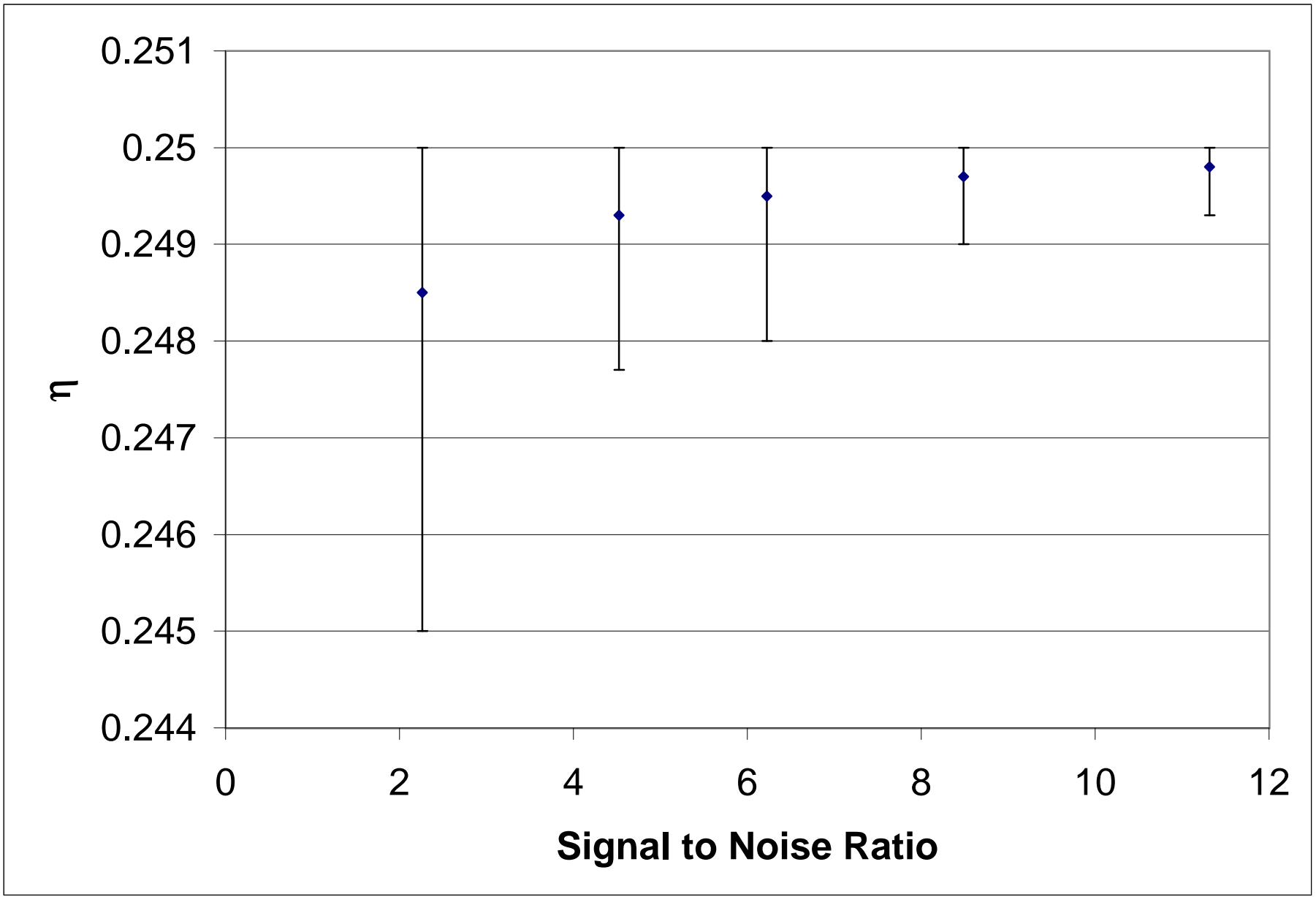
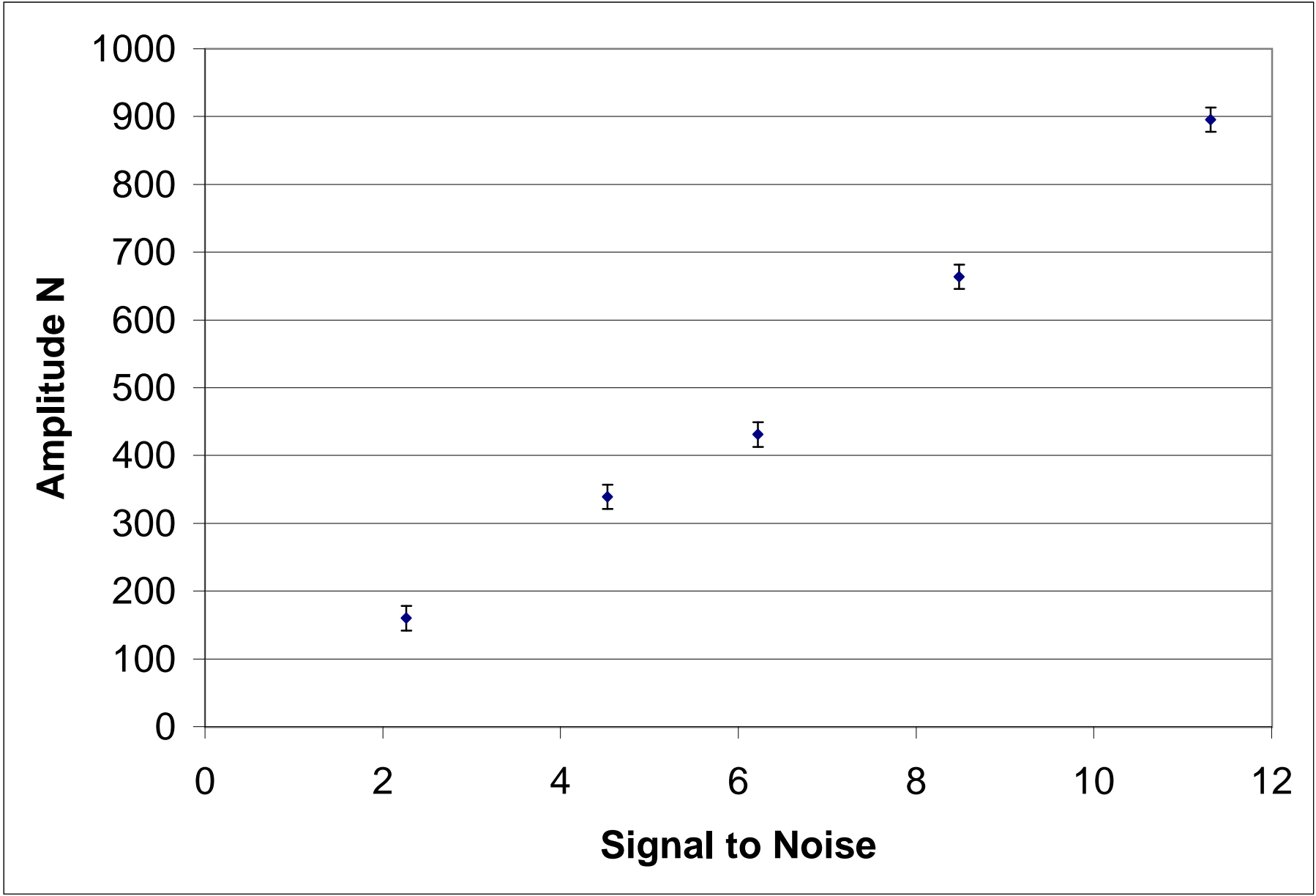
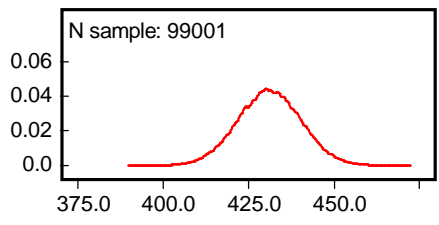
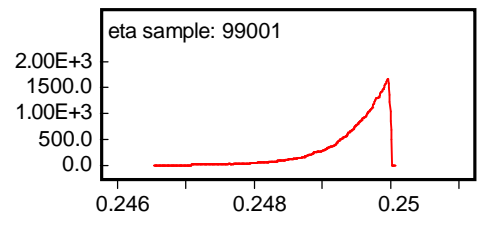
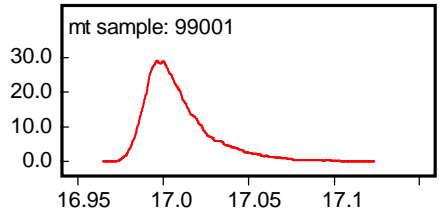
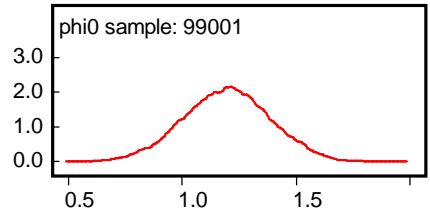
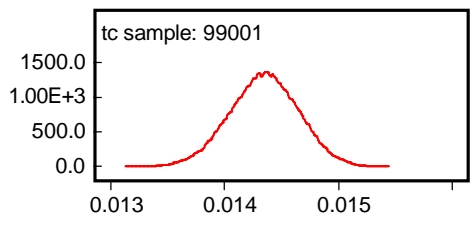
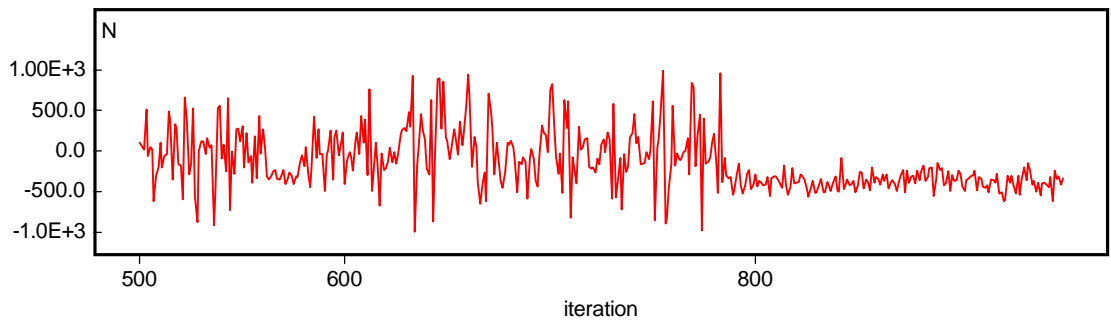
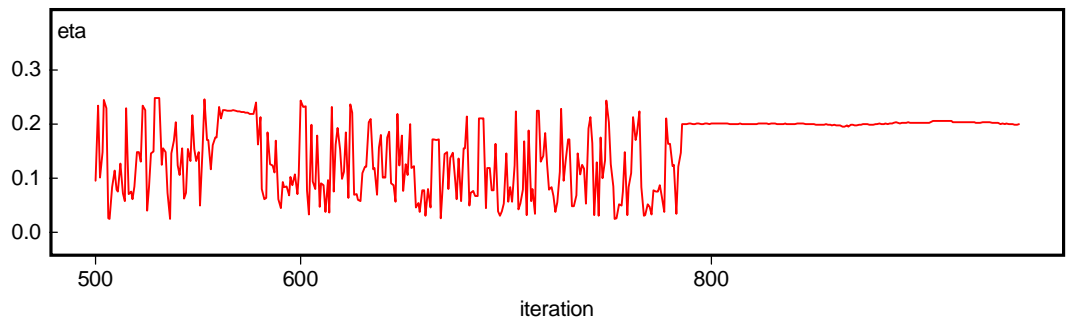
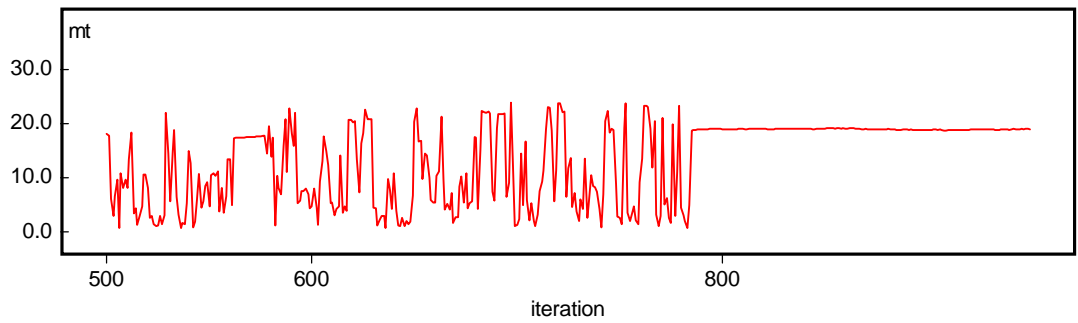
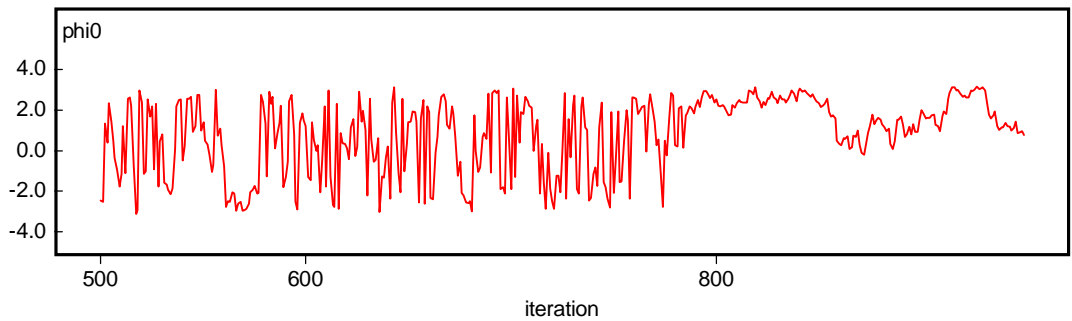
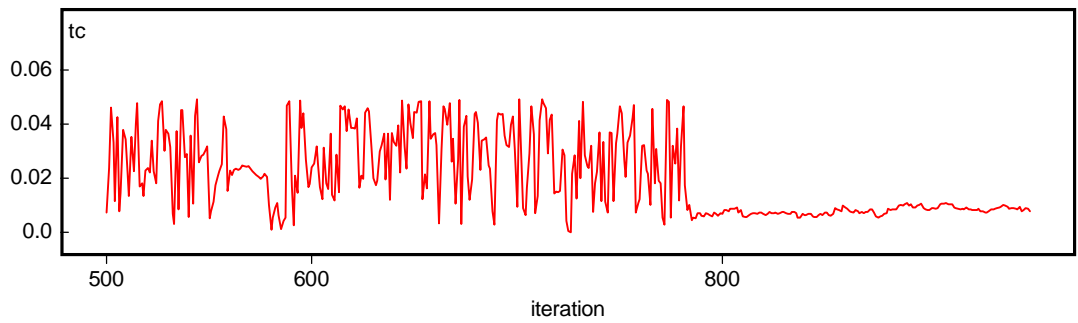
\end{document}